\documentclass{elsart}
\usepackage{graphicx}
\usepackage{amssymb,amsmath}

\newcommand{\be}{\begin{equation}}
\newcommand{\ee}{\end{equation}}

\newcommand{\dd}{{\rm d}}
\newcommand{\vv}[1]{{\bf #1}}

\newcommand{\xx}{\vv{x}}
\newcommand{\F}{{\sc FiEstAS}}

\newcommand{\nn}{N_{\rm node}}
\newcommand{\spl}{_{\rm split}}

\newcommand{\IS}[1]{\hat{#1}_{\rm IS}}

\newcommand{\IMC}{\hat{I}_{\rm MC}}
\newcommand{\DMC}{\hat{\Delta}_{\rm MC}}

\begin{document}

\begin{frontmatter}



\title{FiEstAS sampling -- a Monte Carlo algorithm for multidimensional numerical integration}


\author{Yago Ascasibar}
\address{Astrophysikalisches Institut Potsdam\\ An der Sternwarte 16, Potsdam D-14482, Germany}
\address{Universidad Aut\'onoma de Madrid\\ Dpto F\'{i}sica Te\'orica, Campus de Cantoblanco, Madrid E-28049, Spain}
\ead{yago.ascasibar@uam.es}

\begin{abstract}

This paper describes a new algorithm for Monte Carlo integration, based on the Field Estimator for Arbitrary Spaces (\F).
The algorithm is discussed in detail, and its performance is evaluated in the context of Bayesian analysis, with emphasis on multimodal distributions with strong parameter degeneracies.
Source code is available upon request.

\end{abstract}

\begin{keyword}
numerical integration \sep Bayesian inference \sep Monte Carlo importance sampling

\PACS 02.60.Jh \sep 02.50.Tt \sep 02.70.Uu
\end{keyword}
\end{frontmatter}

  \section{Introduction}
  \label{secIntro}

Many problems in Physics, as well as in other branches of science, involve the computation of an integral
\be
I=\int_V f(\xx)\ \dd\xx
\label{eqI}
\ee
over a given multidimensional domain $V$.
In most circumstances, the value of $I$ has to be estimated numerically, evaluating the function $f$ at a series of $N$ sample points $\xx_i$.

Numerical quadrature algorithms (see e.g. \cite{Press+92}) sample $f$ on a regular grid or at a carefully selected set of unequally-spaced points (Gaussian quadrature).
This approach is very efficient for smooth functions, but it is certainly far from optimal when the integrand is strongly peaked.
Adaptive techniques provide a solution to this problem by refining those sub-intervals of the integration domain where $f$ is found to vary on smaller scales.

In multiple dimensions, Monte Carlo methods usually yield better accuracy for a given $N$ than repeated integrations using one-dimensional deterministic algorithms.
The essence of the Monte Carlo method is that the sample points $\xx_i$ are chosen (to some extent) at random.
The most basic version uses a uniform probability distribution over the integration region, and the estimator for the integral $I$ is given by
\be
\IMC = \frac{V}{N}\sum_{i=1}^{N} f_i
\label{eqI0}
\ee
where $V$ denotes the integration volume and $f_i\equiv f(\xx_i)$.
The error associated to $\IMC$ can be estimated as
\be
\DMC^2 = \frac{V^2}{N^2}\sum_{i=1}^{N} \left( f_i - \frac{\IMC}{V} \right)^2 .
\label{eqD0}
\ee

Although the error decreases as $N^{-1/2}$ (this is, in fact, why Monte Carlo methods are preferable in multidimensional problems), the basic Monte Carlo is still computationally inefficient when the function $f$ has a very narrow peak compared to the total integration volume.
On the one hand, a large number of sample points would be required to obtain good estimates $\IMC$ and $\DMC$.
On the other hand, and more important, it is indeed quite possible that the algorithm achieves convergence, reporting a wrong value of the integral with an extremely small estimate of the error, before finding the maximum (or maxima) of $f$.

A refinement of this method, known as importance sampling, consists in drawing the points from a distribution similar in form to the integrand, so that they are more likely to come from the regions that contribute most to $I$.
Rather than sampling the whole integration volume uniformly, one chooses a probability distribution $g(\xx)$ that adapts to the shape of the integrand.
For any $g(\xx)$, the integral~(\ref{eqI}) can be re-written as
\be
I=\int_V \frac{f(\xx)}{g(\xx)}~g(\xx)~\dd\xx
\ee
and the estimators of $I$ and its error become
\be
\IS{I} = \frac{1}{N}\sum_{i=1}^{N} \frac{f_i}{g_i}
\label{eqI_IS}
\ee
and
\be
\IS{\Delta}^2 = \frac{1}{N^2}\sum_{i=1}^{N} \left( \frac{f_i}{g_i} - \IS{I} \right)^{\!2}
\label{eqD_IS}
\ee
respectively.
The key issue is, of course, the choice of the distribution $g(\xx)$.
These two expressions reduce to~(\ref{eqI0}) and~(\ref{eqD0}) for the particular case of uniform sampling, i.e. all $g_i\equiv g(\xx_i)=1/V$.
In the ideal case, when $g(\xx)=|f(\xx)|/I$, all points contribute exactly the same amount to the sum in~(\ref{eqI_IS}) and zero in~(\ref{eqD_IS}), and the algorithm takes only a few $N$ to achieve the desired accuracy.
Unfortunately, choosing $g(\xx)$ is not straightforward when the shape of $f$ is unknown \emph{a priori}, and evaluating it is actually part of our goal.

A perfect example of such a situation is Bayesian inference, where one tries to select the model $M$ that best describes the data $\vv{D}$ and estimate the values of the model parameters $\vv{\Theta}$.
For a given model, Bayes' Theorem states that
\be
p(\vv{\Theta}|M,\vv{D}) = \frac{ p(\vv{\Theta}|M)~p(\vv{D}|M,\vv{\Theta}) }{ p(\vv{D}|M) }
\label{eqP}
\ee
where $p(\vv{\Theta}|M)$ is the prior probability distribution of the model parameters, $p(\vv{D}|M,\vv{\Theta})$ is the likelihood of the data, $p(\vv{\Theta}|M,\vv{D})$ is the posterior probability distribution of the parameters, and $p(\vv{D}|M)$ is the evidence for model $M$,
\be
p(\vv{D}|M) = \int p(\vv{\Theta}|M)~p(\vv{D}|M,\vv{\Theta})~\dd\vv{\Theta}
\label{eqE}
\ee
The posterior probability distribution~(\ref{eqP}) contains all the information needed for parameter estimation, and the evidence plays in this case the role of a mere normalization constant.
Nevertheless, the value of the integral~(\ref{eqE}) becomes critical in model selection.
Applying again Bayes' Theorem, two models $M_1$ and $M_2$ can be compared by evaluating the Bayes' factor,
\be
\frac{ p(M_1|\vv{D}) }{ p(M_2|\vv{D}) } =
\frac{ p(M_1)~p(\vv{D}|M_1) }{ p(M_2)~p(\vv{D}|M_2) }
\ee
where $p(M_i)$ denotes the prior probability of each model, $p(\vv{D}|M_i)$ are their evidences, and $p(M_i|\vv{D})$ are the posterior probabilities.
Even if one is only interested in parameter estimation, it is good practice to quote the evidence (as well as the assumed priors) in order to make possible the comparison with future work.

The computational part of a Bayesian analysis consists thus in the evaluation of the integral $I=p(\vv{D}|M)=\int f(\vv{\Theta})~\dd\vv{\Theta}$, where $f(\vv{\Theta})= p(\vv{\Theta}|M)~p(\vv{D}|M,\vv{\Theta})$, over the region of the parameter space where $p(\vv{\Theta}|M)>0$.
Since each evaluation of the likelihood may involve a significant computational cost, the algorithm should locate the the maxima of $f(\vv{\Theta})$ as efficiently as possible, even when this function has a complicated structure.
In particular, it is not uncommon that $f(\vv{\Theta})$ presents several maxima, corresponding to different sets of parameter values that provide a good fit to the data.
Which solution is to be preferred depends, within the Bayesian framework, on the value of the integral (the evidence) over the region associated to each peak.
Even more often, there are strong degeneracies in parameter space.
The maxima of the function $f$ can be extremely elongated, generally with a curved shape arising from a non-linear relationship between two or more parameters.

These two problems (multimodality and curved degeneracies) have proven difficult to overcome for many standard algorithms, especially as the number of dimensions increases.
Markov Chain Monte Carlo methods (see e.g. \cite{MacKay03}) typically require a relatively large number of evaluations.
Nested sampling \cite{Skilling04} provides a more efficient alternative, transforming the multidimensional integral~(\ref{eqI}) into a single dimension by means of a re-parameterization in terms of the `prior mass' $p(\vv{\Theta}|M)~\dd\vv{\Theta}$.
This formalism has recently been extended to cope with the problems of multimodality and curved degeneracies by resorting to clustering algorithms \cite{Shaw+07,FerozHobson_07}.
On the other hand, traditional importance sampling methods (e.g. Vegas \cite{Lepage78}) assume a separable weight function, i.e. $g(\xx)=\prod_{d=1}^Dg_d(x_d)$, which can give a good approximation to $f(\xx)$ only as long as the characteristic features are well aligned with the coordinate axes.
An adaptive refinement technique has been implemented in the algorithm Suave \cite{Hahn05} to improve the performance in the general case.

Here I present a new method, based on the Field Estimator for Arbitrary Spaces (\F) developed by \cite{AscasibarBinney05} to estimate the continuous density field underlying a given point distribution.
The algorithm is fully described in Section~\ref{secMethod}, results are presented in Section~\ref{secResults}, and the main conclusions are summarized in Section~\ref{secConclus}.

  \section{Description of the algorithm}
  \label{secMethod}

\F\ sampling performs multidimensional numerical integration by means of Monte Carlo importance sampling.
Here I discuss in detail the most relevant algorithmic issues; actual source code is available upon request.

The method has three free parameters: the desired relative accuracy $\epsilon$, the fraction of points that are uniformly distributed $\eta_u$, and the fractional increase in the number of points per step $\eta_n$.
Default values are $\epsilon=0.01$, $\eta_u=0.1$, and $\eta_n=0.1$.
For the first step, the number of new sample points is set to $n=1$.

The main loop of the program can be summarized as
\begin{enumerate}
\item generate $\eta_un$ new points uniformly
\item compute $g(\xx)$ for this step using \F
\item sample $n_F=(1-\eta_u)n$ new points from $g(\xx)$
\item evaluate the integral and its error
\item increase $n$ by a fraction $\eta_n$
\item repeat until the desired accuracy is achieved
\end{enumerate}

\subsection{Uniform sampling}

\F\ sampling is intended to adapt the probability distribution $g(\xx)$ to the shape of the integrand within the fewest possible number of iterations.
As will be shown below, the presence of pronounced curved degeneracies does not pose a significant problem to the algorithm, but the discovery of narrow, isolated maxima is, at best, a very hard problem, and a large number of sample points may be required in order to achieve the correct result.
Even worse, there can be \emph{no} guarantee that \emph{all} the maxima of $f$ have been located, and therefore it is important to reach a compromise between the detection of potential multimodality and sampling efficiency.

Sampling from a uniform distribution is obviously the most unbiased way to explore the integration domain in search of previously unknown maxima.
The fraction of points (i.e. computational effort) devoted to such exploration is controlled by the parameter $0\le\eta_u\le1$.
A low value would be adequate for functions where multimodality is not a concern, either because the positions of the maxima are already known by other means, because there can only be one, or because the different peaks are not too isolated (i.e. the contrast between the maximum and the saddle point towards the nearest detected peak is not large).
A high value of $\eta_u$ would help the algorithm locate the peaks, but it would render it slow.
What constitutes a reasonable compromise between accuracy and computational resources depends, of course, on the details of the application being considered, and the adopted default value should not preclude the user from applying his/her own judgment.

\subsection{Choice of $g(\xx)$}

The prescription adopted for $g(\xx)$ is arguably the heart of the method.
The idea is to use the set of sample points computed so far to estimate the shape of the integrand.
First, each point is assigned a hypercubical cell by the \F\ algorithm (see \cite{AscasibarBinney05} and Appendix~\ref{secF}).
Let the volumes of these cells be denoted as $v_i$.
Then,
\be
g(\xx) =\frac{\sum_{i=1}^{N} w_i\,\Pi_i(\xx)}{\sum_{i=1}^{N} w_i\,v_i}
\label{eqG}
\ee
with $\Pi_i(\xx)=1$ if $\xx$ belongs to the $i$-th cell, and 0 otherwise.

The choice of weights $w_i=1$ corresponds to uniform sampling, and $w_i=|f_i|$ corresponds to $g(\xx)\approx|f(\xx)|/I$.
This is the optimal probability distribution when the integrand shape is already well described by the sample points, i.e. $f(\xx)\approx\sum_{i=1}^{N} f_i\,\Pi_i(\xx)$.
In practice, convergence is faster if one uses instead
\be
w_i^2 = \left<f^2\right>_i = \frac{1}{n_{\rm nei}}\sum_{j=1}^{n_{\rm nei}} f_j^2
\ee
where the sum is performed over the $n_{\rm nei}$ neighbouring cells in the \F\ tessellation, including the cell $i$ itself.
This prescription tends to sample regions where $|f|$ is large and/or varies on small scales (comparable to the sampling density), and therefore it combines in a simple way the advantages of importance and stratified sampling.
Considering the values of $f$ in adjacent cells is important in order to efficiently explore those regions where large gradients are found, such as newly discovered maxima or pronounced degeneracies.

\subsection{Sampling from $g(\xx)$}

Each step, $n_F=(1-\eta_u)n$ new points are generated from the distribution~(\ref{eqG}).
The probability that each of these points belongs to cell $i$ is given by
\be
p_i = g_i v_i = \frac{w_iv_i}{\sum_{j=1}^{N} w_j\,v_j}
\ee
and therefore the cumulative distribution can be computed as
\be
P_i = \sum_{j=1}^{i} p_j = \frac{P_{i-1}+ w_i\,v_i}{P_N}
\ee
with $P_0=0$.

Sampling from $P_i$ is then trivial.
A uniform random number between 0 and 1 is generated, and the value of $i$ is obtained from a binary search.
The new point is created at a random location, uniformly distributed within cell $i$.

\subsection{Evaluation of the result}

From the $n_F=(1-\eta_u)n$ new points generated during the current step,
\be
\hat{I}_s = \frac{1}{n_F}\sum_{i=1}^{n_F} \frac{f_i}{g_i}
\label{eqIs}
\ee
and
\be
\hat{\Delta}^2_s = \frac{1}{n_F^2}\sum_{i=1}^{n_F} \left( \frac{f_i}{g_i} - \hat{I}_s \right)^{\!2}
\label{eqDs}.
\ee

To estimate the value of $I$, the estimates of the last $S$ steps are combined according to
\be
\hat{I} = \frac{1}{S}\sum_{i=0}^{S-1} \hat{I}_{s-i}
\label{eqItot}
\ee
with an associated error
\be
\hat{\Delta}^2 = \frac{1}{S(S-1)}\sum_{i=0}^{S-1} \left( \hat{I}_{s-i} - \hat{I}\,\right)^{\!2}
\label{eqDtot}.
\ee
The number of steps $S$ is set by the condition
\be
\left|\hat{I}_{s-S}-\hat{I}\,\right|>3\min\left\{\Delta_s,\Delta_S\right\}.
\ee
if $S\ge4$.
Else, the procedure is assumed not to have converged yet, and only the information from the last step is considered (i.e. $\hat{I}=\hat{I}_s$ and $\hat{\Delta}=\hat{\Delta}_s$).

\subsection{Iteration}

At the end of each step, the total number of new points $n$ is increased by a fixed fraction $\eta_n$.
Since the \F\ tree has to be updated once per step in order to estimate $g(\xx)$, a small value of $\eta_n$ increases the overhead of the algorithm.
However, a frequent update results in a more accurate sampling, and therefore a smaller number of function evaluations.
A compromise should be sought for each particular problem, taking into account the computational cost of evaluating the integrand and setting $\eta_n$ so as to minimize the total CPU time.
Otherwise, the precise value of this parameter does not significantly affect the results.

\subsection{Convergence}

The main loop continues generating new points until the estimated relative error drops below the value of the tolerance parameter $\epsilon$,
\be
\hat{\Delta}<\epsilon|\hat{I}|.
\ee
Again, the default value should be interpreted as a guideline, subject to trial and error, common sense, and the specific requirements of the problem at hand.
Note that, as mentioned above, the condition $S\ge4$ is also imposed in order to ensure that the algorithm has actually converged.
This criterion is only important in those few cases where a new maximum has just been discovered at the time $\hat{\Delta}$ decreased below the requested tolerance.

  \section{Results}
  \label{secResults}

The performance of \F\ sampling has been tested by applying the algorithm to a series of toy multidimensional problems with known analytical solution.
The integrands have been chosen to emphasize the issues of multimodality and the presence of non-linear degeneracies.
I also consider the test suite proposed by Genz \cite{Genz86} and compare the results with those of other algorithms.

Two different aspects have been considered: the accuracy of the solution (the actual value of the integral $\hat{I}$, the quality of the error estimate $\hat{\Delta}$, and the ability to find all maxima) and the computational cost (in terms of the number of evaluations of the function $f$ as well as the overhead in CPU time incurred by the \F\ tree).

All the tests have been carried out on a Pentium~4 processor with a clock rate of $3.2$~GHz.
The parameters of the algorithm are always set to their default values.

\subsection{Toy problem I}

As a first example, the method is used to compute the integral of a linear combination of five multivariate Gaussians with different values of $\sigma$, located at arbitrary points in the $xy$-plane.
The locations and dispersions are the same as in \cite{FerozHobson_07}, i.e. $x_i=\{-0.4, -0.35, -0.2,  0.1, 0.45\}$, $y_i=\{-0.4,  0.2,  0.15, -0.15, 0.1\}$, and $\sigma_i=\{0.01,0.01,0.02,0.03,0.05\}$, but each of the Gaussian components has been normalized in order to facilitate the counting of the number of peaks detected,
\be
f(\xx) = \sum_{i=1}^5 \frac{1}{(2\pi\sigma_i^2)^{D/2}} \exp\left[-\frac{|\xx-\vv{c_i}|^2}{2\sigma_i^2}\right]
\ee
where $D$ is the dimensionality of the problem and $\vv{c_i}$ is the centre of each Gaussian, given by $x_i$ and $y_i$.

The correct value of the integral (performed from $-1$ to $1$ along all axes) is thus $I \simeq 5$.
In the Bayesian framework, each of the maxima would correspond to a different solution.
Since all the individual evidences are equal, these five solutions would be equally valid, and none of them would be preferred over the others.
From a frequentist point of view, the narrowest peaks should be favoured because they yield the highest values of the likelihood $f$.

\begin{table}
  \centering
  \begin{tabular}{rrrr}
    \hline
    $\hat{I}$~~ & $\hat{\Delta}$~~ & $N$~~ & $t_{\rm CPU}$ \\
    \hline
    3.982 & 0.037 & 3974 & 0.31 \\
    3.979 & 0.026 & 3603 & 0.27 \\
    5.095 & 0.047 & 3603 & 0.27 \\
    4.026 & 0.024 & 1990 & 0.14 \\
    3.958 & 0.032 & 3603 & 0.27 \\
    4.035 & 0.040 & 3603 & 0.27 \\
    4.032 & 0.039 & 2682 & 0.19 \\
    5.052 & 0.044 & 5325 & 0.42 \\
    4.144 & 0.039 & 3603 & 0.27 \\
    4.998 & 0.048 & 7127 & 0.57 \\
    \hline\\[-4mm]
  \end{tabular}
  \caption
{
    Results obtained by 10 independent runs of \F\ sampling for the toy problem I in two dimensions.
    First and second columns quote the estimates given by equations~(\ref{eqItot}) and~(\ref{eqDtot}), respectively, followed by the total number of evaluations and the CPU time (in seconds) required by the algorithm.
}
  \label{tabToy1a}
\end{table}

\begin{figure}
  \centering \includegraphics[width=8cm]{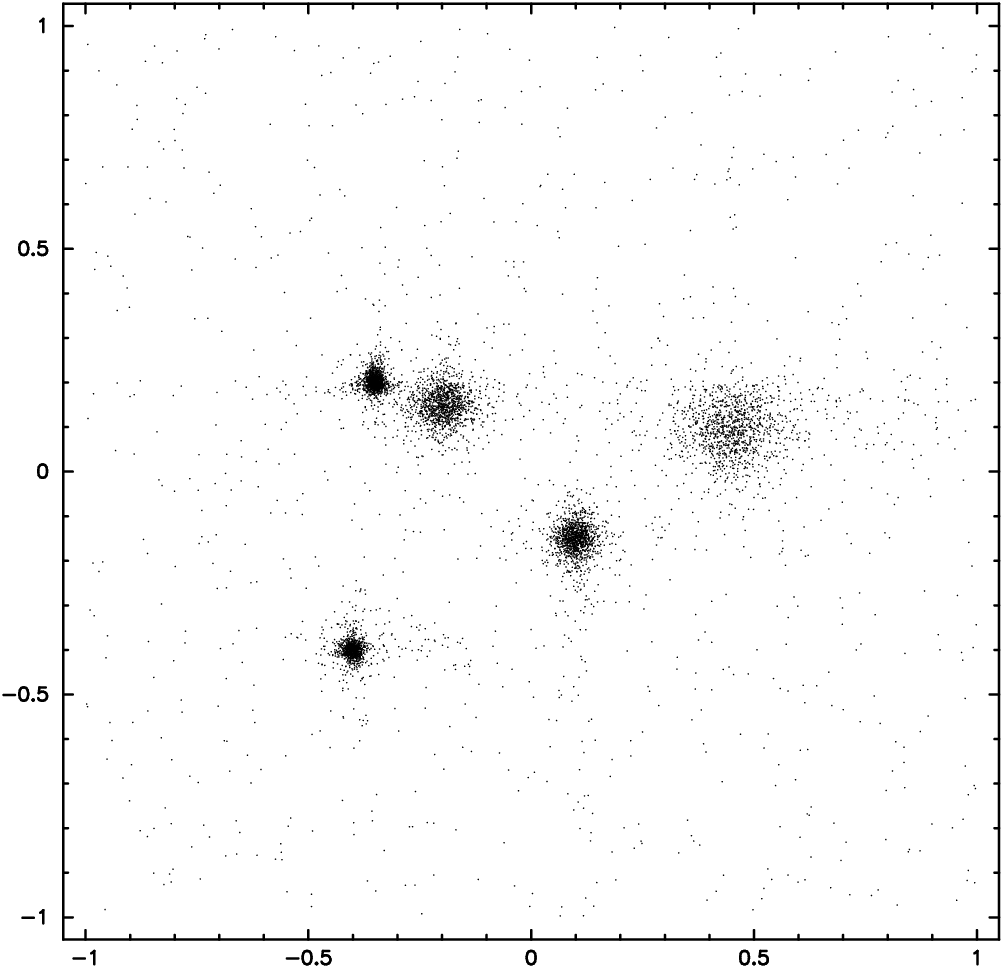}
  \caption
{
    Sample points used by the last run in Table~\ref{tabToy1a}
}
  \label{figToy1a}
\end{figure}

The results of ten independent runs with different random seeds are listed in Table~\ref{tabToy1a} for the two-dimensional case.
In a larger set of one hundred runs, the number of peaks identified by the algorithm was three (4 times), four (56 times), and five (40 times).
The narrow Gaussian at $(-0.4,-0.4)$ was the most difficult to detect.
Although it has the same dispersion as the second component at $(-0.35,0.2)$, the latter is relatively close to another broader maximum, and thus it was much easier to identify.
Neglecting the contribution from missed peaks, the estimated error always corresponded fairly well to the actual deviation with respect to the analytical solution.
The number of sample points required to achieve convergence ranged from $N=1990$ to $11557$, and the run time was in all cases of the order of a few tenths of a second.
The sample points used by the last run are plotted in Figure~\ref{figToy1a}.

\begin{table}
  \centering
  \begin{tabular}{ccccc}
    \hline
    $D$ & $\left<\hat{I}\right>\pm\sigma_I$ & $\left<\hat{\Delta}/\Delta_{\rm true}\right>\pm\sigma_\Delta$ & $\left<N\right>\pm\sigma_N$ & $\left<t_{\rm CPU}\right>\pm\sigma_t$ \\
    \hline
    2 & $4.490\pm0.485$ & $0.767\pm1.836$ & $   4400\pm   1397$ & $    0.34\pm   0.12$ \\
    3 & $3.910\pm0.309$ & $2.128\pm3.254$ & $  18896\pm   6003$ & $    2.87\pm   1.00$ \\
    4 & $4.161\pm0.325$ & $0.722\pm3.254$ & $  86084\pm  40075$ & $   24.64\pm  12.07$ \\
    5 & $3.525\pm1.362$ & $1.441\pm2.796$ & $ 213956\pm 115194$ & $  104.41\pm  59.64$ \\
    6 & $1.790\pm1.577$ & $0.796\pm2.186$ & $ 482739\pm 545036$ & $  380.49\pm 487.6$ \\
    7 & $2.001\pm1.615$ & $1.343\pm2.383$ & $2937754\pm3908586$ & $ 4207.78\pm6332$ \\
    8 & $1.501\pm1.094$ & $0.421\pm4.539$ & $6578407\pm5835573$ & $13554.28\pm15750$ \\
    \hline\\[-4mm]
  \end{tabular}
  \caption
  {
    Results for toy problem I in $D$ dimensions, averaged over ten independent runs.
    Columns show $D$, the value of $\hat{I}$, the deviation of the estimated error $\hat{\Delta}$ with respect to $\Delta_{\rm true}$ (see text), the number of evaluations, and the CPU time in seconds.
  }
  \label{tabToy1b}
\end{table}

The dependence on the number of dimensions $D$ is shown in Table~\ref{tabToy1b}.
The five Gaussians are kept at the same locations in the $xy$-plane, with the same dispersions and normalized to unit evidence, but they refer now to $D$ independent variables ranging from $-1$ to $1$.
Each entry on the table corresponds to the average and standard deviation of ten independent runs with different random seeds.

\F\ sampling typically discovers four or five peaks for $D=2$, which leads to the average value $\left<\hat{I}\right>=4.5\pm0.5$.
As the number of dimensions increases, detecting each of the isolated maxima becomes more difficult, and the value of  $\left<\hat{I}\right>$ decreases.
Note, however, that large dispersions $\sigma_I$ indicate that the number of peaks detected may vary considerably between different runs.
For instance, in $D=6$ dimensions, the algorithm stopped after finding only the broad maximum at $(0.45,0.1)$ in eight out of the ten runs, whereas all peaks were correctly identified in the other two.

The values of $\left<\hat{\Delta}/\Delta_{\rm true}\right>$ correspond to the geometrical average of the estimated error $\hat{\Delta}$ divided by the `true' error $\Delta_{\rm true}=|\hat{I}-I'|$, where $I'$ is the nearest integer to $\hat{I}$.
More precisely,
\be
\left<\hat{\Delta}/\Delta_{\rm true}\right> \equiv
\exp
\left[
  \left< \ln\left( \frac{\hat{\Delta}}{|\hat{I}-I'|} \right) \right>
\right]
\ee
and
\be
\sigma_\Delta \equiv
\exp
\left[
  \left< \ln^2\left( \frac{\hat{\Delta}}{|\hat{I}-I'|} \right) \right> -
  \left< \ln  \left( \frac{\hat{\Delta}}{|\hat{I}-I'|} \right) \right>^2
\right].
\ee
Since all of the Gaussians have unit evidence, $I'$ corresponds to both the number of peaks detected and the analytical value of the integral after subtracting the contribution from the undetected components.
The results obtained, $\left<\hat{\Delta}/\Delta_{\rm true}\right>\sim1\pm3$, suggest that $\hat{\Delta}$ is typically within a factor of three (above or below) from $\Delta_{\rm true}$ (see e.g. Table~\ref{tabToy1a}).

Finally, the exact number of evaluations required depends on the integrand being considered, the requested tolerance $\epsilon$, and, to a lesser extent, the values of the parameters $\eta_u$ and $\eta_n$.
What can be inferred from the data in Table~~\ref{tabToy1b} is that $N$ grows exponentially with the number $D$ of dimensions.
Since the complexity of the \F\ tree is ${\cal O}(N\log N)$ \cite{AscasibarBinney05}, the overhead in CPU time also grows exponentially.

\subsection{Toy problem II}

In order to assess the effect of non-linear degeneracies, let us also investigate the toy problem proposed by \cite{AllanachLester_07},
\be
f(\xx) = R(\xx;\vv{c_1},r_1,w_1) + R(\xx;\vv{c_2},r_2,w_2)
\ee
with
\be
R(\xx;\vv{c},r,w) = \frac{1}{\kappa}\exp\left[-\frac{(|\xx-\vv{c}| -r)^2}{2w^2}\right] .
\ee
The function $R(\vv{x})$ represents a ring of radius $r$ and width $w$, centered at $\vv{c}$.
The normalization $\kappa$ is chosen so that $\int R(\xx) \dd\xx=1$, and it can be expressed as
\be
\kappa = \frac{D\,\pi^{D/2}}{\Gamma(D/2+1)\sqrt{2\pi w^2}} K_{D-1}(r,w)
\ee
where $\Gamma$ denotes the Gamma function and $K_D$ is given by the recursive formula
\be
K_D(r,w) = r\,K_{D-1} + (D-1)w^2\,K_{D-2}
\ee
with $K_0=1$ and $K_1=r$.
The two rings are separated by $7$ units along the $x$-axis, and their radii and widths are set to the values $r_1=1$, $r_2=2$, and $w_1=w_2=0.1$.
The domain of integration ranges from $-6$ to $6$ in all dimensions.

\begin{figure}
  \centering \includegraphics[width=8cm]{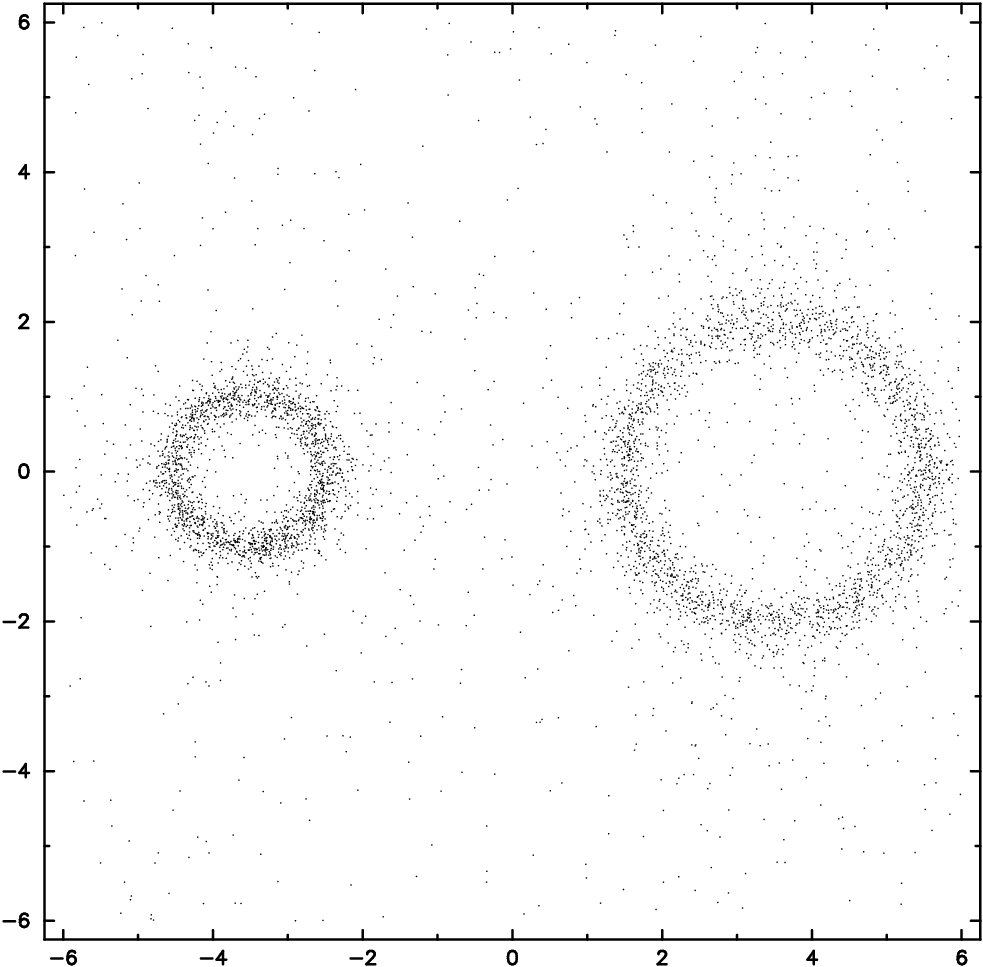}
  \caption
{
    Sample points for toy problem II in two dimensions.
}
  \label{figToy2}
\end{figure}

\begin{table}
  \centering
  \begin{tabular}{ccccc}
    \hline
    $D$ & $\left<\hat{I}\right>\pm\sigma_I$ & $\left<\hat{\Delta}/\Delta_{\rm true}\right>\pm\sigma_\Delta$ & $\left<N\right>\pm\sigma_N$ & $\left<t_{\rm CPU}\right>\pm\sigma_t$ \\
    \hline
    2 & $1.990\pm0.023$ & $1.792\pm2.991$ & $   6127\pm   1012$ & $    0.47\pm   0.09$ \\
    3 & $2.001\pm0.017$ & $1.411\pm2.137$ & $  29660\pm   8082$ & $    4.60\pm   1.35$ \\
    4 & $1.984\pm0.027$ & $0.938\pm2.700$ & $  72619\pm  13452$ & $   20.03\pm   4.04$ \\
    5 & $1.897\pm0.304$ & $1.199\pm2.473$ & $ 179333\pm  56814$ & $   83.86\pm  29.15$ \\
    6 & $1.491\pm0.501$ & $0.933\pm3.181$ & $ 327824\pm  77328$ & $  249.65\pm  64.26$ \\
    7 & $1.196\pm0.397$ & $0.782\pm2.524$ & $ 735387\pm 203382$ & $  552.04\pm 169.9$ \\
    8 & $1.381\pm0.480$ & $0.675\pm2.206$ & $4517358\pm3616613$ & $11001.25\pm9788$ \\
    \hline\\[-4mm]
  \end{tabular}
  \caption
{
    Results for toy problem II.
}
  \label{tabToy2}
\end{table}

The values of the integral computed by \F\ sampling are given in Table~\ref{tabToy2}, and Figure~\ref{figToy2} shows the sample points used by the algorithm in the two-dimensional case.
The two rings have been correctly identified by the algorithm ten times out of ten up to $D=4$.
As the number of dimensions increases, the small ring occupies an exponentially smaller fraction of the integration domain with respect to the bigger structure, and the detection rate decreases.
For $D=8$, it is found in four of the ten runs.

The shape of the rings is always accurately recovered, which proves that the presence of strong degeneracies does not significantly affect the accuracy of the method.
As in the previous examples, the estimated error is within a factor of three from the true value, and both the number $N$ of evaluations and the overhead $t_{\rm CPU}$ grow exponentially with $D$.

\subsection{Comparison with other algorithms}

The method presented here significantly outperforms the standard Monte Carlo with uniform probability, and its results are comparable to those of the bank sampling technique described in \cite{AllanachLester_07}, at least for problems of low dimensionality.
An important advantage of \F\ sampling is that prior information on the shape of $f$ is not required, although it could be easily taken into account by simply adding the `bank' points at any stage.
On the other hand, comparison with the results of \cite{FerozHobson_07} suggests that clustered nested sampling may be more efficient for large $D$.

\begin{table}
  \centering
  \begin{tabular}{crrrrrr}
    \hline
    Model & $D$
    & $N_{\rm Vegas}$ & $N_{\rm Suave}$
    & $N_{\rm Divonne}$ & $N_{\rm Cuhre}$
    & $N_{\rm FiEstAS}$
    \\ \hline I
    & 2 &   52000 &   90000 &   18569 &      9165 &     3974 \\
    & 3 &   45000 &  160000 &  193910 &     82677 &    14012 \\
    & 4 &  175000 &  320010 &  285312 &    215271 &   168641 \\
    & 5 &  314500 & 1270259 &   86567 &   1150695 &   204091 \\
    & 6 &       - &  120036 &  171526 &   3971451 &   224517 \\
    & 7 &       - &  280084 &  236217 &  26245785 & 11186912 \\
    & 8 &       - & 1840397 & 6619086 & 116242685 &  7640740 \\
    \hline II
    & 2 &   45000 &   60000 &    6321 &    19175 &    5325 \\
    & 3 &   67500 &   90000 &   24700 &   213233 &   33236 \\
    & 4 &   85000 &  230006 &   99300 &  8675253 &   58999 \\
    & 5 &  137500 &  320015 &  189236 & 96301023 &  328797 \\
    & 6 &       - &  510046 &  219956 &        - &  397883 \\
    & 7 &       - &  820128 &  364261 &        - &  640912 \\
    & 8 &       - & 1120153 &  122915 &        - & 1828970 \\
    \hline\\[-4mm]
  \end{tabular}
  \caption
  {
    Number of evaluations required by the algorithms Vegas \cite{Lepage78}, Suave \cite{Hahn05}, Divonne \cite{FriedmanWright81}, and Cuhre \cite{Berntsen+91} for toy problems I and II in $D$ dimensions.
    Empty entries indicate failure to converge within $N=10^9$ function calls.
    Results of a single run of \F\ sampling are quoted in the last column.
  }
  \label{tabCompN}
\end{table}

\begin{table}
  \centering
  \begin{tabular}{crrrrrr}
    \hline
    Model & $D$
    & $\hat{I}_{\rm Vegas}$ & $\hat{I}_{\rm Suave}$
    & $\hat{I}_{\rm Divonne}$ & $\hat{I}_{\rm Cuhre}$
    & $\hat{I}_{\rm FiEstAS}$
    \\ \hline I
    & 2 & 5.01 & 5.00 & 4.95 & 5.00 & 4.06 \\
    & 3 & 2.29 & 3.99 & 5.00 & 5.00 & 4.09 \\
    & 4 & 1.03 & 3.99 & 3.64 & 2.00 & 4.56 \\
    & 5 & 2.11 & 3.95 & 2.61 & 2.00 & 3.96 \\
    & 6 &    - & 0.99 & 2.01 & 2.00 & 1.00 \\
    & 7 &    - & 0.37 & 2.47 & 2.00 & 4.99 \\
    & 8 &    - & 1.80 & 0.58 & 1.99 & 3.01 \\
    \hline II
    & 2 & 2.00 & 2.00 & 2.02 & 2.00 & 2.00 \\
    & 3 & 2.00 & 2.00 & 1.96 & 1.63 & 2.01 \\
    & 4 & 1.95 & 2.00 & 2.01 & 1.54 & 1.98 \\
    & 5 & 1.37 & 2.00 & 1.38 & 1.60 & 2.00 \\
    & 6 &    - & 1.26 & 1.20 &    - & 2.01 \\
    & 7 &    - & 0.50 & 1.00 &    - & 0.96 \\
    & 8 &    - & 0.30 & 0.51 &    - & 0.99 \\
    \hline\\[-4mm]
  \end{tabular}
  \caption
  {
    Estimates $\hat{I}$ returned by the different algorithms (same runs as in Table~\ref{tabCompN}).
  }
  \label{tabCompI}
\end{table}

Tables~\ref{tabCompN} and~\ref{tabCompI} show the performance of the algorithms Vegas \cite{Lepage78}, Suave \cite{Hahn05}, Divonne \cite{FriedmanWright81}, and Cuhre \cite{Berntsen+91}, as implemented in the Cuba library\footnote{http://www.feynarts.de/cuba} \cite{Hahn05}, when applied to toy problems~I and~II.
Table~\ref{tabCompN} quotes the number of integrand evaluations, and Table~\ref{tabCompI} shows the estimated value of the integral.
\F\ sampling is quite competitive in terms of efficiency; the values of $N$ are always comparable to or lower than those required by the other methods.
Its main drawback, however, is the computational overhead; the CPU times in Tables~\ref{tabToy1b} and~\ref{tabToy2} are much larger than those required by the other algorithms (a few minutes in the worst cases).
In terms of accuracy, the number of detected maxima compares very favourably with the other methods.
Moreover, the estimate $\hat{I}$ is almost always close to the true value of the integral, ignoring the contribution from undetected peaks; among all the examples quoted, only the run for toy model~I in four dimensions reported a wrong value.

In order to probe a wider variety of integrands, the different methods are also benchmarked with the Genz test suite \cite{Genz86}, based on the following six function families:
\begin{enumerate}
\item Oscillatory: $$f_1(\xx)=\cos(\vv{c}\cdot\xx+2\pi w_1)$$
\item Product peak: $$f_2(\xx)=\prod_{d=1}^{D}\frac{1}{(x_i-w_i)^2-c_i^{-2}}$$
\item Corner peak: $$f_3(\xx)=(1+\vv{c}\cdot\xx)^{-(D+1)}$$
\item Gaussian: $$f_4(\xx)=\exp(-|\vv{c}\cdot(\xx-\vv{w})|^2)$$
\item $C^0$-continuous: $$f_5(\xx)=\exp(-\vv{c}\cdot|\xx-\vv{w}|)$$
\item Discontinuous:
  $$f_6(\xx)=
  \begin{cases}
    \exp(\vv{c}\cdot\xx) & \text{if $x_1<w_1$ and $x_2<w_2$}\\
    0 & \text{otherwise}
  \end{cases}$$
\end{enumerate}
The parameter vector $\vv{c}$ sets the overall difficulty of the problem.
Its components are chosen as uniform random numbers from 0 to 1, and then they are normalized according to
\be
||\vv{c}||_1 \equiv \sum_{d=1}^D c_d = 50
\ee
except for the oscillatory integrand, for which I set $||\vv{c}||_1 = 5$.
The components of $\vv{w}$ are uniform random numbers between 0 and 1.
In general terms, they do not affect the difficulty of the problem, since they only control the location of the features (e.g. maxima) of $f$.
They play an important role, though, for the oscillatory family, because $w_1$ sets the precise value of the integral, which can be arbitrarily close to zero.
This poses a problem for \F\ sampling (and, more generally, to pure Monte Carlo algorithms) because a large number of evaluations is required to achieve the cancellation of positive and negative terms with the desired accuracy.

\begin{table}
  \centering
  \begin{tabular}{crrrrrr}
    \hline
    Genz & $D$
    & $\left<N_{\rm Vegas}\right>$
    & $\left<N_{\rm Suave}\right>$
    & $\left<N_{\rm Divonne}\right>$
    & $\left<N_{\rm Cuhre}\right>$
    & $\left<N_{\rm FiEstAS}\right>$
    \\
    \hline 1
    & 2 &$^{(2)}$97056 &  66500 &        2954 &          195 &$^{(3)}$165913 \\
    & 5 &        18350 &  42000 &        2917 &          819 & $^{(1)}$22110 \\
    & 8 &        68500 &  51500 &        3697 &         3315 & $^{(1)}$57411 \\
    \hline 2
    & 2 &         5625 &  20500 &          3186 &           2190 &   1096 \\
    & 5 &         9100 &  31500 &         10624 &         477422 &  16559 \\
    & 8 &        12300 &  40000 &         23326 &$^{(19)}$469625 &  57301 \\
    \hline 3
    & 2 &         4500 &  13500 &          1690 &            195 &    395 \\
    & 5 &         4500 &  20000 &          3970 &            819 &   3231 \\
    & 8 &         7000 &  20000 &          5495 &           3315 &  11505 \\
    \hline 4
    & 2 &         5750 &  23000 &          7137 &           1768 &   2037 \\
    & 5 &        16950 &  42000 &         30582 &          57630 &  32559 \\
    & 8 &        30200 &  64000 & $^{(3)}$52943 &$^{(10)}$549185 & 183580 \\
    \hline 5
    & 2 &         5375 &  22500 &          4542 &           4407 &   1595 \\
    & 5 &        10025 &  33500 &         19747 &$^{(17)}$671853 &  21697 \\
    & 8 &        16650 &  42500 & $^{(1)}$40161 &$^{(20)}$   -   &  92247 \\
    \hline 6
    & 2 &         7825 &  32500 &          6994 &          50453 &   3056 \\
    & 5 &        18025 &  87008 & $^{(1)}$64257 &          84766 &  25940 \\
    & 8 &$^{(1)}$36079 & 103006 &$^{(4)}$152156 &$^{(12)}$604435 & 115577 \\
    \hline\\[-4mm]
  \end{tabular}
  \caption
  {
    Average number of evaluations required by each algorithm for the test suite proposed by Genz \cite{Genz86}.
    The small numbers in parentheses indicate the number of runs where the desired accuracy was not reached for $N=10^6$.
  }
  \label{tabGenz}
\end{table}

Results of the Genz test in 2, 5, and 8 dimensions are quoted in Table~\ref{tabGenz} for the different algorithms.
Each entry shows the average number of evaluations over 20 random realizations of each family.
In order to speed up the test, we impose a maximum number of evaluations $N_{\rm max}=10^6$.
It is clear from the comparison that every algorithm has its own strengths and weaknesses.
Some methods are better suited to solve a particular class of problems, and some others perform better in many dimensions.
In general terms, the results presented here suggest that \F\ sampling could be a good choice for complicated integrands of low to moderate dimensionality.

  \section{Conclusions}
  \label{secConclus}

This paper presents a new algorithm to carry out numerical integration in multiple dimensions or, in a Bayesian context, sample from the posterior distribution and compute the evidence for the assumed model.
The method, dubbed `\F\ sampling', is a variant of importance sampling where the weight of each point is computed with the help of the Field Estimator for Arbitrary Spaces \cite{AscasibarBinney05}.

Its performance has been tested for several toy problems with known analytical solution, specifically designed to contain multimodal distributions and significant degeneracies.
The results suggest that \F\ sampling provides an interesting alternative to other methods for problems of low dimensionality.
In particular, it is able to discover most isolated maxima of the integrand, and it can perfectly sample from distributions with pronounced degeneracies.
As the number of dimensions increases, the ability of the algorithm to identify peaks decreases and the required computational resources (in terms of both time and memory) grow exponentially.


 \appendix

  \section{\F}
  \label{secF}

The Field Estimator for Arbitrary Spaces (\F) is a technique to estimate the continuous (probability) density field underlying a given distribution of data points.
Particular attention is paid to avoid imposing a metric on the data space.
Indeed, the problem may actually be regarded as \emph{computing} the appropriate metric, given the data.

\F\ assigns each point a volume $v_i$ by means of a $k-d$ tree.
The space is recursively divided, one dimension at a time, until there is only one single data point in each node.
The original implementation, described in \cite{AscasibarBinney05}, is heavily oriented towards a particular problem, namely the estimation of densities in phase space (a non-Euclidean, six-dimensional space composed of three-dimensional positions and velocities).
Here I use a more general version of the algorithm, where the dimension less likely to arise from a uniform distribution is selected for splitting at each step (very similar to the Shannon entropy approach followed by \cite{SharmaSteinmetz06}).
More precisely, a histogram with $B=1+\sqrt{\nn}$ bins is built for each dimension, and the log-likelihood
\be
L_d = \ln(\nn!) - \nn\ln(B) - \sum_{b=1}^{B} \ln(n_{bd}!)
\ee
is computed, where the indices $1\le d\le D$ and $1\le b\le B$ denote the dimension and the bin number, respectively, $n_{bd}$ is the number of points in each bin, and $\nn$ is the total number of points in the node.
In order to encourage a similar number of divisions along all dimensions, I add to $L_d$ the number of times $s_d$ that the $d$-th axis has already been split,
\be
L_d' = L_d + s_d .
\ee
The dimension with smaller $L'$ is divided at the point $x\spl=(x_{\rm l}+x_{\rm r})/2$, where $x_{\rm l}$ is the maximum $x$ of all points lying on the 'left' side ($b\le b\spl$) and $x_{\rm r}$ is the minimum $x$ of the points lying on the 'right' ($b>b\spl$) side.
The bin $1\le b\spl<B$ is chosen in order that the number of points on each side is as close as possible to $\nn/2$.

\begin{figure*}
  \includegraphics[width=.23\textwidth]{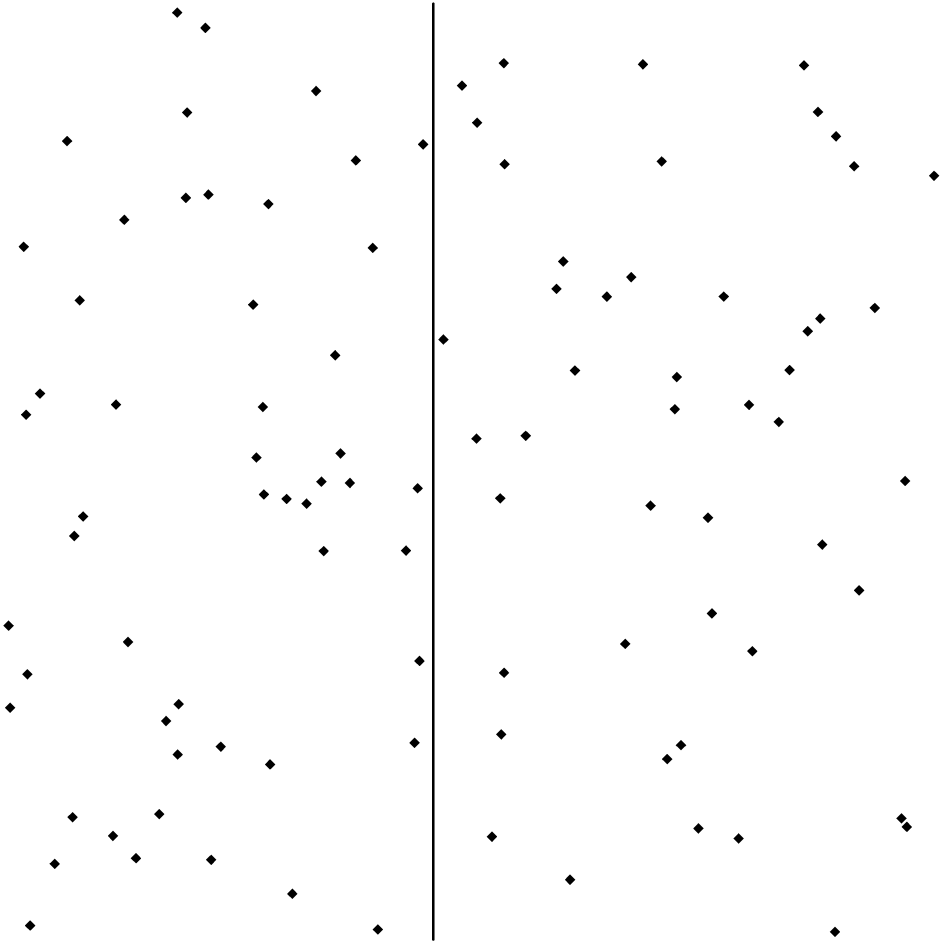}\hfill
  \includegraphics[width=.23\textwidth]{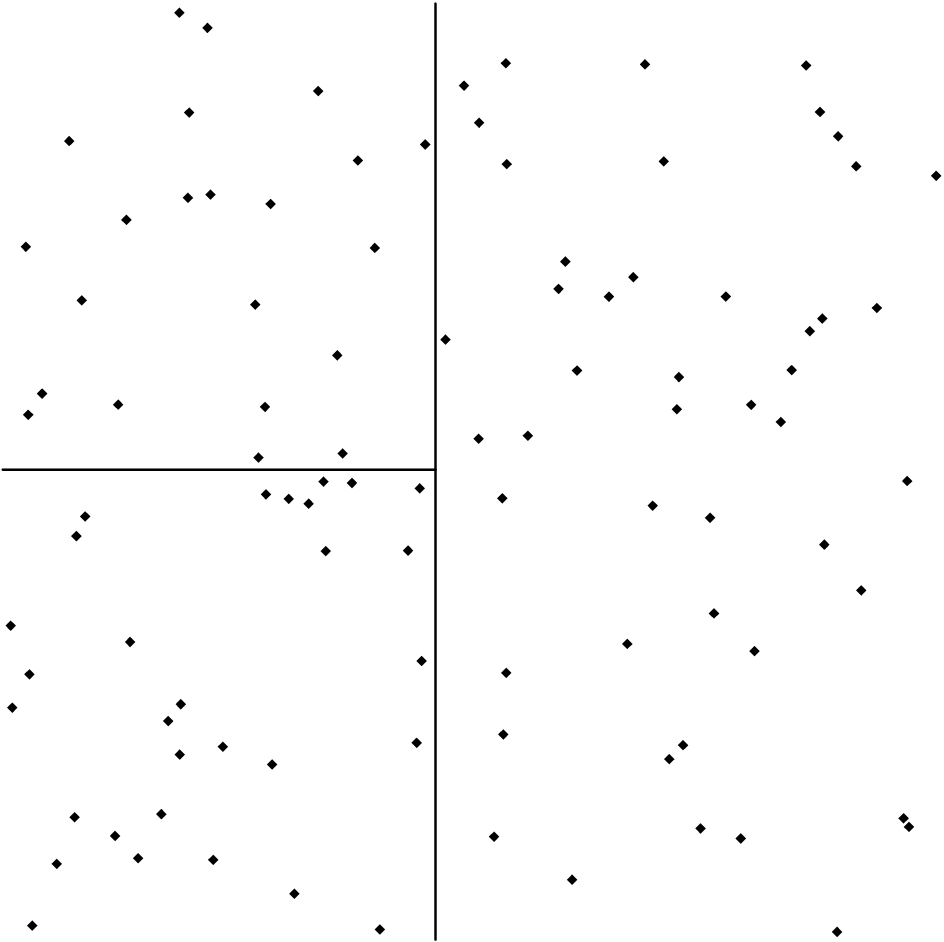}\hfill
  \includegraphics[width=.23\textwidth]{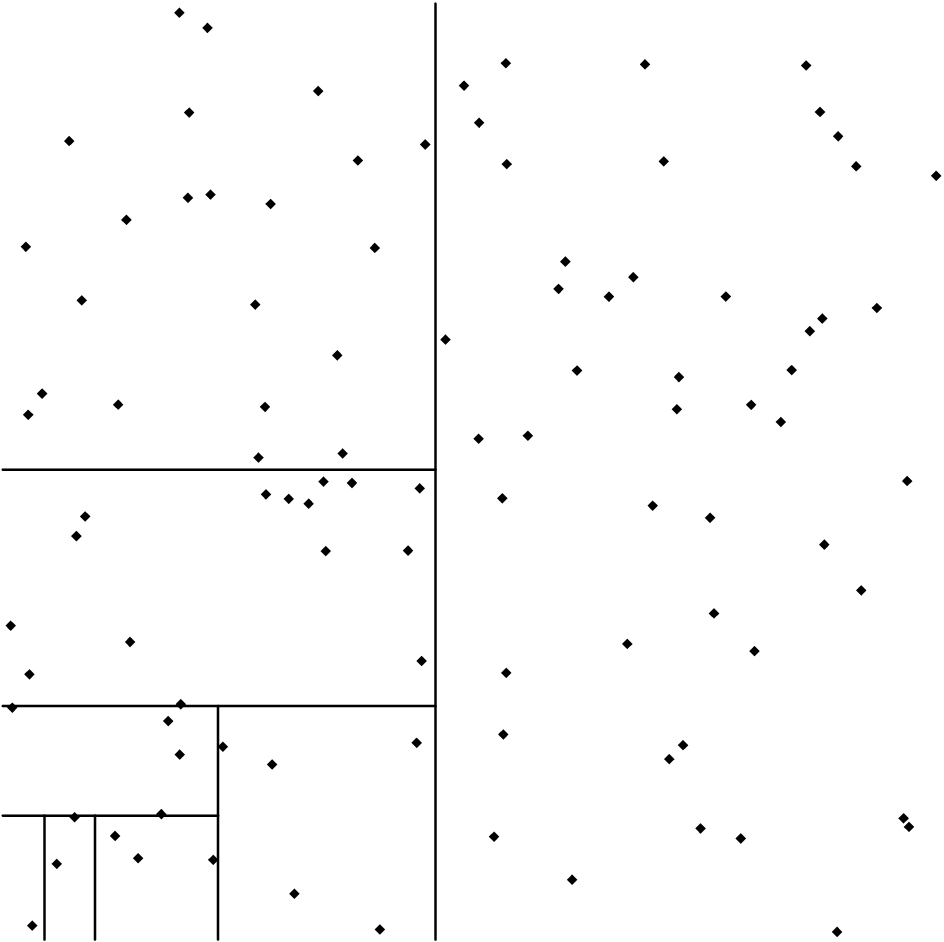}\hfill
  \includegraphics[width=.23\textwidth]{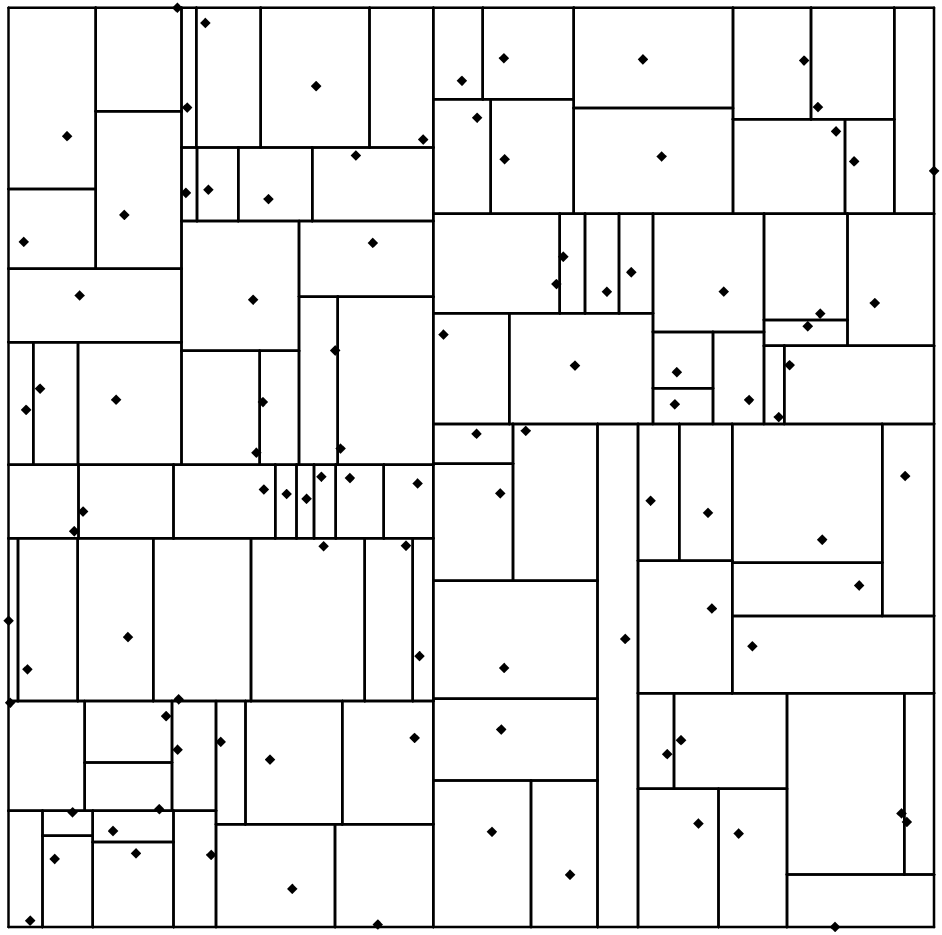}
  \caption
  {
    \F\ algorithm applied to a random uniform distribution of 100 points in two dimensions.
  }
  \label{figF}
\end{figure*}

The procedure is illustrated in Figure~\ref{figF}, where successive steps of the algorithm are plotted for a random realization of 100 uniformly-distributed data points in two dimensions.
For this particular realization, the dimension with smaller $L'$ turned out to be the $x$-axis (first panel in the figure).
Then, for the points on the left side, the $y$-axis was selected for splitting (second panel).
The same axis was chosen again for the bottom region, and the sequence continued by dividing further along the $x$-, $y$-, $x$-, and $x$-axes.
At this moment (third panel in the figure) there is only one point on the left side, and control returns to the parent node in order to split the right branch (containing, in this case, two points).
The final tessellation obtained by this method is shown on the last panel.


 \bibliographystyle{cpc}
 \bibliography{bibliography}


\end{document}